\documentclass[galaxies,article,accept,oneauthors,dvi2pdf]{mdpi} 

\firstpage{1} 
\pubvolume{xx}
\issuenum{1}
\articlenumber{5}
\pubyear{2020}
\copyrightyear{2020}
\history{Received: date; Accepted: date; Published: date}

\Title{X-ray Observations of Planetary Nebulae since WORKPLANS~I and Beyond}

\Author{Mart\'\i n A.\ Guerrero $^1$\orcidA{}}

\AuthorNames{Martin A Guerrero}

\address{%
  $^1$ \quad Instituto de Astrof\'\i sica de Andaluc\'\i a, IAA-CSIC,
  Glorieta de la Astronom\'\i a s/n, E-18008 Granada, Spain; mar@iaa.es}

\corres{Correspondence: mar@iaa.es}

\abstract{
Planetary nebulae (PNe) were expected to be filled with hot pressurized gas
driving their expansion.
\emph{ROSAT} hinted at the presence of diffuse X-ray emission from
these hot bubbles and detected the first sources of hard X-ray
emission from their central stars, but it was not until the advent
of \emph{Chandra} and \emph{XMM-Newton} that we became able to study
in detail their occurrence and physical properties.
Here I review the progress in the X-ray observations of PNe since
the first WORKshop for PLAnetary Nebulae observationS (WORKPLANS)
and present the perspective for future X-ray missions with particular
emphasis on \emph{eROSITA}.
}


\begin{document}


\section{Introduction}

Planetary nebulae (PNe) are the descendants of low- and intermediate-mass
stars, as they eject their envelopes at the tip of the asymptotic giant
branch (AGB) and start their evolution towards the white dwarf phase.
At this time, the central star of the PN (CSPN) develops a fast (although
tenuous) stellar wind that impinges and snowplows the material previously
ejected during the AGB as a slow and dense stellar wind.
In this interacting stellar winds (ISW) model for the formation
and evolution of PNe, the fast stellar wind is shock-heated up
to temperatures that depend on the CSPN terminal wind velocity
as $T\sim v_{\infty}^2$ \citep{Dyson1997}.
For typical CSPN wind velocities of $v_{\infty} = 500-4000$~km~s$^{-1}$
\citep[e.g.,][]{Guerrero2013}, shocked plasma at X-ray-emitting temperatures
in excess of 10$^6$~K can be expected.

The very first \emph{Chandra} observations of PNe already found hot 
bubbles with temperatures of a few million K confined within the 
inner rims of BD$+$30$^{\circ}$3639 \citep{Kastner2000} and NGC\,6543 
\cite{Chu2001}.
This general behavior has been largely confirmed by the \emph{Chandra}
Planetary Nebula Survey \citep[ChanPlaNS;][]{Kastner2012,Freeman2014},
a volume-limited survey of a sample of nearby ($d<1.5$~kpc) PNe.
Diffuse X-ray emission is detected only in compact ($r<0.2$~pc)
relatively young PNe with a closed-shell morphology \citep[as
suggested previously by][]{Ruiz2011}, for a detection rate 
$\sim$30\%.

\emph{Chandra} has also unveiled unexpected sources of hard X-ray emission
($>$0.5 keV) from CSPNe \citep{Guerrero2001}, and ChanPlaNS has confirmed
the prevalence of these hard X-ray point-sources in a significant
number of CSPNe \citep{Montez2015}. 
Some of these hard X-ray sources can be attributed to the coronal 
emission from a dwarf or late-type giant companion \citep{Montez2010}, 
while others can be assigned to shocks in their fast winds as in OB 
and Wolf-Rayet stars, especially among CSPNe with powerful fast stellar 
winds \citep{Guerrero2001}. 
The X-ray to bolometric luminosity ratios of the latter, 
$L_{\rm X}/L_{\rm  bol} \sim 10^{-7}$, are indeed similar to 
those of single OB stars \citep[e.g.,][]{Nebot2018}.

\section{Available X-ray Observatories}

\emph{ROSAT} obtained pointed or serendipituous observations of 63 PNe, 
but only 16 were detected and most of these detections correspond to 
soft photospheric emission from hot CSPNe, whereas only three, namely 
BD$+$30$^{\circ}$3639, NGC\,6543, and NGC\,7009 could be tentatively 
attributed to diffuse hot gas \citep{Guerrero2000}.  
Most modern X-ray observations of PNe have been obtained using the 
\emph{Chandra} and \emph{XMM-Newton} observatories, which were launched 
in 1999, i.e., more than 20 years ago!
Still, they have been used rather than other more modern X-ray observatories
such as \emph{Swift} or \emph{AstroSat} because of the much larger effective
areas of \emph{Chandra} and \emph{XMM-Newton} in the energy range of interest
($\leq$1.5 keV) for PNe.  
Specifically, the effective area of the \emph{Chandra} ACIS-S7 and 
\emph{XMM-Newton} EPIC-pn detectors at 1.5 keV are 5--8 and 9--13 times 
larger, respectively, than those of \emph{Swift} XRT ($\simeq$135 cm$^2$) 
and \emph{AstroSat} ($\simeq$90 cm$^2$).   

\begin{figure}[t]
\centering
\includegraphics[bb=18 220 600 640,width=0.75\columnwidth]{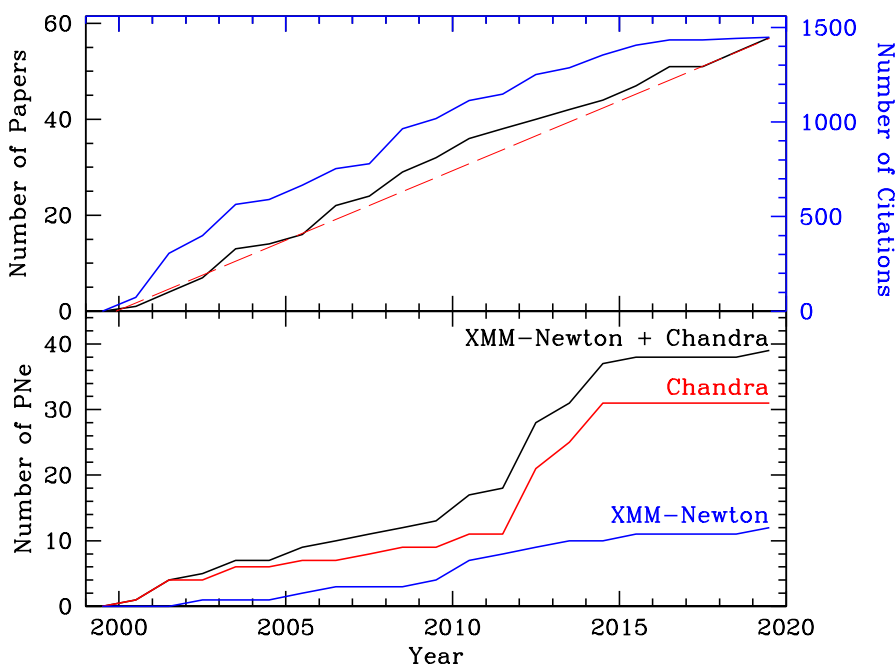}
\caption{
(bottom) 
Time evolution of the number of PNe detected by the \emph{Chandra} (red) 
and \emph{XMM-Newton} (blue) X-ray observatories.  
(top) 
Time evolution of the number of papers (black) and citations to these papers 
(blue) based on those X-ray observations (source Astronomycal Database System,
ADS, at https://ui.adsabs.harvard.edu).  
Note that the left and right axes in this panel refer to numbers of 
papers and citations, respectively, as denoted by their labels and 
colors.  
The red-dashed line is a linear fit to the number of papers, implying 
a publication rate $\simeq$3 papers per year on the X-ray emission
from PNe.  
}
\label{fig.stat}
\end{figure}   

Since \emph{Chandra} and \emph{XMM-Newton} started operations in 1999, 
the number of PNe with diffuse X-ray emission and CSPNe with hard X-ray 
emission from grown steadily.  
As illustrated in the bottom panel of Figure~\ref{fig.stat}, 
\emph{Chandra} and \emph{XMM-Newton} have revealed these types 
of X-ray emission in almost 40 PNe.  
The notable increase of PNe between 2010 and 2015 can be attributed to 
the effort of the PN community to obtain the ChanPlaNS survey, which has 
resulted in unprecedented progress in the understanding of the nature 
of point and extended sources of X-rays in PNe.  
The impact of all these observations has been tremendous.  
Almost 60 papers have been written in this topic since the launch of 
\emph{Chandra} and \emph{XMM-Newton} at a publication rate of three 
papers per year (top panel of Figure~\ref{fig.stat}).  
By the end of 2019, these papers had garnered close to 1,500 citations.  

\begin{figure}[t]
\centering
\includegraphics[bb=25 230 585 535,width=0.75\columnwidth]{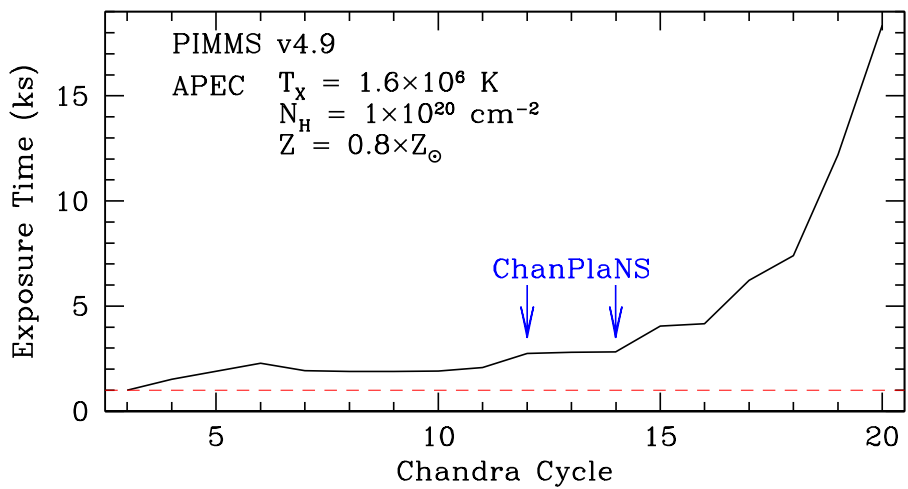}
\caption{
Time evolution of the exposure time required for a \emph{Chandra} ACIS-S 
observation to obtain a given number of counts for a plasma model typical 
of the diffuse X-ray emission in PNe.  
Simulations were obtained using the \emph{Chandra} PIMMS v4.9 
simulator, which includes the most up-to-date calibrations.  
A plasma with a typical 1.6$\times$10$^6$ K temperature, 
low hydrogen column density, and slightly subsolar chemical
composition has been assumed.  
}
\label{fig.acis}
\end{figure}

The bottom panel of Figure~\ref{fig.stat} reveals a worrisome fact.  
Despite its unbeatable spatial resolution, \emph{Chandra} has not 
contributed to any new X-ray detection among PNe since 2014 
(\emph{XMM-Newton} has made two new discoveries since then).  
This can be mostly attributed to 
the continuous degradation of \emph{Chandra} ACIS's sensitivity to low
energy photons, which has made it more and more unlikely to detect the
soft X-ray emission from PNe with sufficient photon counts for detailed
spectral analysis\footnote{
  The sensitivity of \emph{Chandra} HRC has remained basically unchanged
  in this time period, but the spectral capability of this instrument,
  which is otherwise very well suited for high-resolution imaging studies,
  is very limited.}.  
This is illustrated in Figure~\ref{fig.acis}, showing the time evolution 
of the exposure time required for a \emph{Chandra} ACIS-S observation to 
obtain the same number of counts for a plasma model typical of the diffuse 
X-ray emission in PNe.  
Similarly, the simulation of the \emph{Chandra} ACIS-S count number and 
image of BD+30$^\circ$3639 through Cycle 10 ($\simeq$2500 cnts), Cycle 15 
($\simeq$1500 cnts) and Cycle 20 ($\simeq$350 cnts) compares very poorly 
to those observed in Cycle 1 ($\simeq$4400 cnts, Montez, priv.\ 
communication). 
In this sense, the ChanPlaNS survey was very timely, 
using the last useful \emph{Chandra} window to acquire
imaging spectroscopic observations of PNe.  

\section{Latest Results on X-ray Emission from PNe}

Since the WORPLANS~I meeting in December 2015, a number of new results 
based on X-ray observations or theoretical investigations have been 
reported.  
These can be grouped in four different categories.  

\subsection{Theoretical Studies}

The large amount of observational data already available has allowed the 
refinement of our ISW models on the production of X-ray-emitting gas in 
PNe.  
Two different research lines have been pursued.  
The special properties of the hot bubbles produced by CSPNe with H-deficient 
stellar winds has been extensively investigated, with particular attention 
paid to the effects of heat conduction \citep{Sandin2016} and its application 
to the case of BD+30$^\circ$3639 \citep{Heller2018}.  
On the other hand, the combined effects of heat conduction and turbulent 
mixing have been tested using high-resolution, two-dimensional, 
radiation-hydrodynamical simulations \citep{Toala2016,Toala2018}.  
In these detailed models, instabilities in the wind-wind interaction zone 
produce clumps and filaments in the swept-up shell of nebular material 
with notable effects in the time-dependent X-ray emission as the stellar 
parameters change.  
It is shown that the diffuse X-ray emission at early times is dominated 
by the contribution from the hot, shocked stellar wind, whereas the 
contribution from nebular gas dominates at later times.  
Furthermore, these models demonstrate that the X-ray temperature of a hot
bubble is anti-correlated with metallicity, but its X-ray luminosity increases 
with metallicity.  
The conclusions are quite robust, turbulent mixing layers are the origin 
of the soft X-ray emission in the majority of diffuse nebulae.

\subsection{Multi-wavelength Studies}

Multi-wavelength studies combining X-ray and infrared, optical, and UV 
observations have provided very interesting results in the past few years.  
The obvious correlation between the X-ray morphology and the varying local 
extinction, as the soft X-ray emission from PNe is easily absorbed by even 
small amounts of material along the line of sight, has been confirmed through 
the comparison of optical and infrared images of BD+30$^\circ$3639 
\citep{FK2016} and NGC\,7027 \citep{MK2018} with archival or new 
\emph{Chandra} observations.  
The discovery of the mixing layer at the interface between the X-ray-emitting
hot bubble and the optical photo-ionized nebular shell in NGC\,6543 using
\emph{HST} STIS observations \citep{Fang2016} is very exciting. 
The emission in the N~{\sc v} $\lambda\lambda$1239,1243 \AA\ lines 
from this interface layer has allowed for the first time not only 
to probe it, but to determine its spatial extent, thickness and 
physical properties.  
These are critical to assess the relative importance of heat conduction and 
turbulent mixing in the production of hot gas in PNe.  

\subsection{X-ray Variability Studies}

The hard X-ray emission from the CSPN of the Eskimo Nebula has been found
to be variable with a period $\simeq$6 hours \citep{Guerrero2019}.  
This X-ray emission might be attributed to accretion of material from the 
CSPN wind onto a white dwarf companion, and thus the observed period can 
be assigned either to the orbital period of the companion or to the rotation 
of an accretion disk around it. 
Time-analysis of the X-ray emission from CSPNe can be 
very revealing of the nature of putative companions.  

\subsection{New Discoveries of X-ray PNe} 

Only one more PN has been found to exhibit diffuse X-ray emission since 
the WORKPLANS~I.  
This is the case of NGC\,5189, a PN around a [WO1] CSPN whose extended 
X-ray emission has been discovered by \emph{XMM-Newton} \citep{Toala2019}.  
The spatial distribution and spectral properties of the X-ray emission, 
particularly its high carbon abundances, are consistent with a born-again 
scenario where a CSPN becomes carbon-rich through a very late thermal pulse 
(VLTP) event and then develops a fast, carbon-rich wind that generates X-ray
emission.  
The physics of the interaction between the present fast stellar 
wind of the CSPN and the hydrogen-poor clumps of material ejected in the 
VLTP event is fascinating \citep[see also the cases of A\,30 and 
A\,78,][]{Guerrero2012,Toala2015}.  

\section{The Future of X-ray Observations of PNe}

In the long run, 
the \emph{X-ray Imaging and Spectroscopic Mission} (\emph{XRISM}), 
the \emph{X-ray Universe Baryon Surveyor} (\emph{HUBS}), 
the \emph{Advanced Telescope for High ENergy Astrophysics} (\emph{Athena}), 
and 
the \emph{Lynx X-ray Observatory} (\emph{Lynx}), 
with their enhanced collecting areas, high-dispersion instruments, and/or 
spatial resolution will play major roles in a better understanding of the 
production of hot gas in the interior of PNe and its role in their evolution 
and shaping.  
Meanwhile, the \emph{eROSITA} mission will produce an X-ray map of the 
whole sky with unprecedented sensitivity.  
The final \emph{eROSITA} All Sky Survey (eRASS) is expected to 
have a sensitivity in the range from 1$\times$10$^{-14}$ down 
to 3$\times$10$^{-17}$ erg~cm$^{-2}$~s$^{-1}$, depending on the 
final exposure time of a particular position on the sky.  

\begin{figure}[t]
\centering
\includegraphics[bb=25 310 585 500,width=0.9\columnwidth]{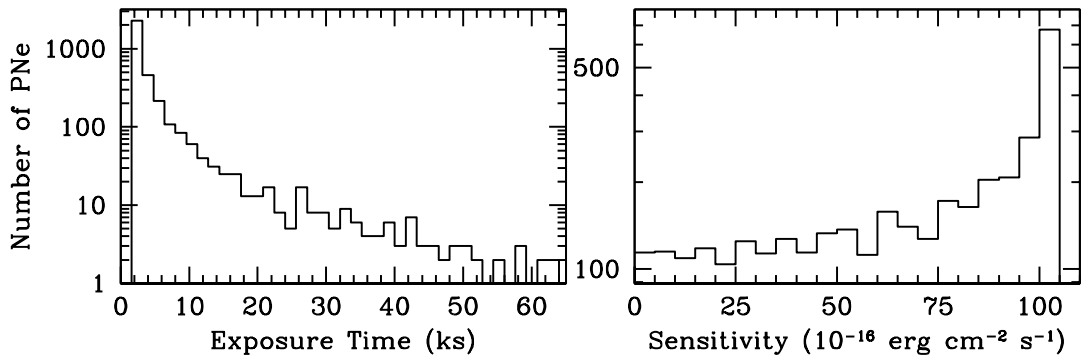}
\caption{
Distribution of exposure time (left) and sensitivity (rigth) for PNe 
expected in the final eRASS map of the whole sky.  
Most PNe will be located in areas of the eRASS with total exposure times 
shorter than 20 ks, but a significant number of them, $\simeq$200, will 
have longer exposure times.  
Accordingly, a significant number of PNe, those with long 
exposure times in the final eRASS, will have observations 
with sensitivities adequate to detect X-ray fluxes smaller 
than $\simeq$2$\times$10$^{-15}$ erg~cm$^{-2}$~s$^{-1}$.  
}
\label{fig.erass}
\end{figure}

We have convolved the expected sensitivity map of the final eRASS with the 
position of all known PNe to determine their exposure times and X-ray 
sensitivity.  
The expected distributions of exposure times and sensitivities for 
the whole sample of known PNe are shown in Figure~\ref{fig.erass}.  
It reveals that $\simeq$200 PNe will be observed with exposure 
times greater than 20 ks.  
As the seven X-ray telescopes of \emph{eROSITA} have a total effective area 
above that of \emph{XMM-Newton} EPIC-pn, almost 500 PNe will have observations 
as sensitive, $\leq$2$\times$10$^{-15}$ erg~cm$^{-2}$~s$^{-1}$, as those of PNe 
in the ChanPlaNS survey. 
Obviously this will produce a major improvement in our statistics of the 
presence of X-ray emission in PNe.  

\section{Conclusions}

Since the advent of the modern X-ray observatories \emph{Chandra} and 
\emph{XMM-Newton}, the number of PNe with diffuse X-ray emission and 
CSPNe with hard X-ray emission have grown steadily.  
The large number and quality of these X-ray data sets of PNe 
has resulted in major advances in a research field for which
sensitive observations basically did not exist before 1999.  
Over 60 papers have been generated 
based on these observations since 1999, 
garnering almost 1,500 citations.  

This publication rate has remained since the last WORKPLANS meeting in
December 2015, but the emphasis of the studies has changed significantly.  
In this time period, notable advances in theoretical analyses have
been made and multi-wavelength and time-domain studies are gaining
momentum, while the discoveries of new cases for X-ray emission from PNe 
or their CSPNe has slowed nearly to a halt, mostly due to the dramatic
decrease of sensitivity to soft X-ray photons of \emph{Chandra} ACIS.  

In the near future, \emph{eROSITA} can make a major contribution 
to our understanding of the production of hot gas in PNe and its 
effects in their evolution and shaping.  
Hundreds of PNe can be expected to be observed with sensitivities 
similar to many of the currently available observations carried 
out by \emph{Chandra} and \emph{XMM-Newton} in the last two decades.

\vspace{6pt} 




\funding{
This research was funded by the Spanish Ministerio de Ciencia, Innovaci\'on 
y Universidades grant number PGC2018-102184-B-I00 co-funded by FEDER funds.  
}

\acknowledgments{
The author acknowledges the scientific discussions with 
Joel H.\ Kastner, Rodolfo Montez, Quentin Parker, and 
Jes\'us A.\ Toal\'a, which helped him assess the impact of 
\emph{eROSITA} and its eRASS for the research of X-ray 
emission from PNe.  
Four anonymous referees are also acknowledged for their useful comments 
and suggestions to improve this manuscript.  
}

\conflictsofinterest{The authors declare no conflict of interest.}

\reftitle{References}

\end{document}